# Simple implementation of deletion from open-address hash table

*Maxim A. Kolosovskiy*
*Altai State Technical University, Russia*
*maxim.astu@gmail.com*

*Abstract*

*Deletion from open-address hash table is not so easy as deletion from chained hash table, because in open-address table we can't simply mark a slot containing deleted key as empty. Search for keys may become incorrect. The classical method to implement deletion is to mark slots in hash table by three values: "free", "busy", "deleted". That method is easy to implement, but there are some disadvantages. In this article we consider alternative method of deletion keys, where we avoid using the mark "deleted". The article contains the implementation of the method in Java.*

# 1. Introduction

*Hash table* is an effective data structure to store dynamic set. Hash tables support following operations with that set:
- addition new element into the set;
- searching element in the set;
- deletion element from the set.

Under some assumptions, these operations can be executed in O(1).

The simplest implementation of such data structure is an ordinary array, where k-th element corresponds to key *k*. Thus, we can execute all operations in O(1). It is impossible to use this implementation, if the total number of keys is large.

We can reduce the amount of used memory, if instead of used key *k* as array index we use value of function *h(k)*. This function is called **hash function**, the value *h(k)* is called **hash value** of key *k*. Using of such functions is the main idea of hash tables. Thus, the size of array equals the number of possible values of hash function. There are many ways to construct a hash function, but in this article we will use very simple function: $h(k) = |k \% m|$, where *m* – size of array.

We've reduced the amount of memory, but have obtained the following problem. If we work with two or more keys, which have the same hash value, these keys map to the same cell in the array. Such situations are called **collisions**. There are two basic ways to implement hash tables to resolve collisions:
- chained hash table;
- open-address hash table.

In chained hash table each cell of the array contains the linked list of elements, which have corresponding hash value. To add (delete, search) element in the set we add (delete, search) to corresponding linked list. Thus, time of execution depends on length of the linked lists.

## 2. Open-address hash tables

In open-address hash table we store all elements in one array and resolve collisions by using other cells in this array. To perform insertion we successively examine some slots in the table, until we find an empty slot or understand that the key is contained in the table. To perform search we execute similar routine.

The sequence for examining slots depends on added (searched) key and method, which we use to determine a next slot to be examined. Denote this sequence in the following way: $h(k) = h(k, 0), h(k, 1), h(k, 2) \ldots$. It is called **probe sequence**. There are several ways to build these sequences:

- linear probing

    $h(k, i) = (h(k) + C\,i)\ \%\ m$, where $C$ is constant.

- quadratic probing

    $h(k, i) = (h(k) + C_1\,i + C_2\,i^2)\ \%\ m$, where $C_1$, $C_2$ are constants.

- double hashing

    $h(k, i) = (h(k) + h'(k)\,i)\ \%\ m$, where $h'(k)$ is second hash function.

In our examples we will use the linear probing with constant $C = 1$. In our implementation we must use the linear probing, but may change the value of constant $C$.

Lets look at the implementation of open-address hash table, which can add and search keys (*a* – array for storing keys, *ex* – array, which signs cells in array *a* as empty or busy)

```
class HashTable {
    final int SIZE = 1000000; // size of the table
    long[] a = new long[SIZE];
    boolean[] ex = new boolean[SIZE];
    // adds new key. Returns false, if key is contained already
    boolean add(long key) {
        int hash = (int) Math.abs(key % SIZE);
        for (; ex[hash]; hash = (hash + 1) % SIZE)
            if (a[hash] == key)
                return false;
        ex[hash] = true;
        a[hash] = key;
        return true;
    }
    // returns true, if key is contained in the table
    boolean contains(long key) {
        int hash = (int) Math.abs(key % SIZE);
        for (; ex[hash]; hash = (hash + 1) % SIZE)
            if (a[hash] == key)
                return true;
        return false;
    }
}
```

## 3. Deletion keys from the table

Unlike chained hash table, open-address table doesn't use any linked lists. But we can imagine, that there are implicit lists, which can contain keys with different hash values (unlike lists in chained tables, where each list contains only keys with the same hash value). Each busy cell in the array is the beginning of the list in open-address table. Busy cell in the array (in open-address table) can be contained in some lists. It is the main obstacle to delete elements from these implicit lists, because we cannot mark deleted slot as empty – it will destroy the lists.

The classical solution for saving these lists in correct state is to mark deleted slot by special value "deleted". We can add new key into this cell, but cannot interrupt a search, if we reach such cell. The disadvantage of this method is too long lists, which we obtain after many deletions, because such deletion doesn't reduce sizes of lists. In the worst case we can have actually empty table (with many slots marked "deleted"), but search may take a lot of time, because we should examine all "deleted" slots in the probe sequence.

In this paper we propose another method to delete keys from open-address hash table. After deletion of key some gaps are appeared in lists, which contain deleted slot. Therefore, we will compress lists, which contain deleted slot, so these lists remain in correct state. Deleting routine divides into two parts:
- search deleted key and mark a slot as empty;
- compress all lists, which include deleted key.

The first part isn't difficult, it is very similar to searching routine (see code for clarifications). Let's consider second part.

After marking deleted slot as empty we continue to pass the probe sequence until reach a free slot. We are keeping index of empty slot, originally it equals index of deleted slot (the variable *free* in our implementation). If the current element in the sequence can be moved to a free cell, we move it and change the variable *free*, because the current slot becomes free.

When we can move the current element to the free cell? We can move the current element, if the free cell and the current element can be in the same list, i.e. when we were adding the current element, we could add it in this free cell, but this cell was busy and we continued searching for a free cell.

To make this decision we use the array *ex* and the variable *off*. *off* is keeping offset from the free cell to the current element. *(ex[i]-1)* equals the amount of busy slots, which we skipped, while were adding key *a[i]* (see the code of method *add*). So the condition to determine to move or not the key *a[i]* is *ex[i] > off*.

Let's look at the final implementation of hash table.

```
class HashTable {
    final int SIZE = 1000000; // size of the table
    long[] a = new long[SIZE];
    int[] ex = new int[SIZE];
    // adds new key
    boolean add(long key) {
        int i = h(key), j = 1;
        for (; ex[i] != 0; i = (i + 1) % SIZE, j++)
            if (a[i] == key)
                return false;
        ex[i] = j;
        a[i] = key;
        return true;
    }
    // returns true, if key is contained in the table
    boolean contains(long key) {
        for (int i = h(key); ex[i] != 0; i = (i + 1) % SIZE)
            if (a[i] == key)
                return true;
        return false;
    }
    // removes key from the table
    boolean remove(long key) {
        for (int i = h(key); ex[i] != 0; i = (i + 1) % SIZE)
            if (a[i] == key) {
                ex[i] = 0;
                compress(i);
                return true;
            }
        return false;
    }
    // compresses lists
    void compress(int free) {
        int i = (free + 1) % SIZE, off = 1;
        for (; ex[i] != 0; i = (i + 1) % SIZE, off++)
            if (ex[i] > off) {
                // move current element
                a[free] = a[i];
                ex[free] = ex[i] - off;
                // mark current slot as free
                ex[i] = 0;
                off = 0;
                free = i;
            }
    }
    int h(long key) {
        return (int) Math.abs(key % SIZE);
    }
}
```

## 4. Conclusion

In this paper we obtain implementation of data structure, which stores dynamic set. It is possible to add, search and remove elements in O(1). We can see, that the number of iterations for deletion equals the number of iterations for addition and searching, because in all these routines we should find a free slot.

The disadvantage of this implementation is obligation to use the linear probing, which can produce primary clustering. So we should choose hash function more attentive.